\documentclass [12pt,twoside] {article}
\usepackage{setspace,graphicx,natbib,bm,fancyhdr}
\usepackage{amsmath,amsthm,amssymb,latexsym}
\usepackage{color}
\usepackage{graphicx}

\usepackage{authblk}

\newcommand{\rf}{\vskip .1in\par\sloppy\hangindent=1pc\hangafter=1
                 \noindent}

\newcommand{\ch}[1]{\mbox{$\stackrel{\sim}{#1}$}}

\newcommand{\slas}[1]{\mbox{${{#1} \!\!\! /}$}}

\begin{document}
\title{\bf An approach to classical quantum field theory based on the geometry of locally conformally flat space-time} 
\author{John Mashford \\
School of Mathematics and Statistics \\
University of Melbourne, Parkville, Vic. 3010, Australia \\
E-mail: mashford@unimelb.edu.au}
\date{\today}
\maketitle

\begin{abstract}

This paper gives an introduction to certain classical physical theories described in the context of locally Minkowskian causal structures (LMCSs). For simplicity of exposition we consider LMCSs which have locally Euclidean topology (i.e. are manifolds) and hence are M\"{o}bius structures. We describe natural principal bundle structures associated with M\"{o}bius structures. Fermion fields are associated with sections of vector bundles associated with the principal bundles while interaction fields (bosons) are associated with endomorphisms of the space of fermion fields.
Classical quantum field theory (the Dirac equation and Maxwell's equations) is obtained by considering representations of the structure group $K\subset U(2,2)$ of a principal bundle associated with a given M\"{o}bius structure where $K$, while being a subset of $U(2,2)$, is also isomorphic to $SL(2,{\bf C})\times U(1)$. The analysis requires the use of an intertwining operator between the action of $K$ on ${\bf R}^4$ and the adjoint action action of $K$ on $u(2,2)$ and it is shown that the Feynman slash operator, in the chiral representation for the Dirac gamma matrices, has this intertwining property.

\end{abstract}

Mathematical Subject Classifications MSC codes:
20G45 Group theory: applications to physics
22E10 General properties and structure of complex Lie groups
51P05 Geometry and Physics
53A30 Conformal differential geometry
\newline \newline

\section{Introduction}

It may be appropriate to use the term ``classical quantum field theory" to signify the physics described by the Dirac equation (which supercedes the Schr\"{o}dinger equation), Maxwell's equations and Einstein's general theory of relativity (GR).

The principal concept of GR is that space-time can be represented as a Lorentzian 4-manifold, $X$ say, and that gravitation is associated with the metric tensor, $g$ say, of $X$. Einstein proposed that $g$ is related to a tensor $T$, the energy-momentum tensor, which is generated by the (non-gravitational) fields of the universe and that this relationship can be described by the Einstein field equations. Thus in Einstein's theory gravitation is associated with the geometry of space-time (described by the metric $g$) and is coupled to the non-gravitational fields which, apart from the condition of Einstein's field equations have an independent existence. 
Thus the ``data" or given mathematical structure of GR is 
\begin{enumerate}
\item a smooth 4-manifold, $X$ say, 
\item  a Lorentzian metric $g$ for $X$,
\item electromagnetic (and possibly other) fields together with their field equations,
\item Einstein's field equations relating the non-gravitational fields to the gravitational field.
\end{enumerate}
Einstein believed that ultimately GR would evolve into a theory in which all physical fields are associated with the geometry of space-time.

The work described in the present paper models space-time in a fashion following from, or at least in the spirit of, GR. Our work focuses on locally conformally flat space-time and derives classical electrodynamics and quantum mechanics in this context.   

It was suggested in [1] that space-time could be
represented as (or has the structure of) an acyclic digraph. Light 
beams were defined as being maximal sets of interacting events. Light beams can
be given the natural topology associated with their total ordering and
space-time can be topologized with the topological union of its light beams. 
Such spaces were called {\em webs} [1] or {\em causal structures} [2].

Minkowski space can be given the structure of a causal structure in a natural
way. Two events are considered to be interacting if they both lie on a common null ray.
The topology for Minkowski space generated by this causal structure is
strictly finer than the Euclidean topology.

Physical space-time is considered to be a causal structure which is locally
isomorphic to Minkowski space. Such spaces will be called locally Minkowskian
causal structures (LMCSs).
If an LMCS is given the atlas of
all causal charts then it is not a manifold. If it is given an atlas such that
all charts in the atlas have ranges which are Euclidean open sets and such that
the overlap isomorphisms are $C^{\infty}$ diffeomorphisms then the LMCS is a
$C^{\infty}$ 4-manifold and will be known as a Euclidean locally Minkowskian
causal structure (ELMCS).

Let $J$ be the group generated by the Lorentz transformations 
and the
positive dilatations. Then $J \cong O(1,3) \times (0,\infty)$. 
A diffeomorphism $f:U\rightarrow V$ for $U$ and $V$ open subsets of
Minkowski space is conformal if and only if
\begin{equation} \label{eq:conformal_transformation}
 (Df)(x) \in J, \forall x \in U.
\end{equation}
Let $\Gamma(1,3)$ be the pseudogroup [3] consisting of
$C^{\infty}$ conformal transformations in Minkowski space.
By a (Lorentzian) M\"{o}bius structure [4] will be meant a set $X$ which has an
atlas of charts
${\cal A} = \{ (U_i,\phi_i) : i \in I \}$ such that 
\begin{eqnarray}
 & & (\forall i \in I)\  U_i \subset X , V_i=\phi_i(U_i) \mbox{ is a (Euclidean)
 open subset of ${\bf R}^4$ and} \nonumber \\
 & & (\forall i,j \in I)\  \phi_i \circ \phi_j^{-1} \in \Gamma(1,3). \nonumber
\end{eqnarray}

Any ELMCS is a M\"{o}bius structure [2]. A M\"{o}bius structure is an ELMCS if and
only if the overlap diffeomorphisms are orthochronous and there are no closed 
polygonal curves for which every
side is a future directed light ray.

Any locally conformally flat Lorentzian manifold can be viewed as a M\"{o}bius 
structure. The map from locally conformally flat Lorentzian manifolds to M\"{o}bius
structures is many to one. However, in the theory described here, 
 it will not be assumed that there is given a distinguished metric. From the physical point of
view this means that the theory described here does not have a metric
as a ``background structure'' but relies on 
an underlying causal structure.

One might think that M\"{o}bius structures do not have a rich enough structure to model the diversity and complexity of the universe from microscopic to macroscopic scales. In fact they do have a rich and variable structure. The simplest (4D) M\"{o}bius structure is ${\bf R}^4$ itself and the next most simple way to construct M\"{o}bius structures is to take an arbitrary open subset of ${\bf R}^4$. Thus, for example, one can consider a M\"{o}bius structure ${\bf R}^4\setminus\bigcup_{i=1}^n C_i$ where the $C_i$ are closed subsets, where the boundary of any set $C_i$ may be connected by a ``wormhole" to the boundary of another such set  within the given copy of ${\bf R}^4$ or to a subset of another copy of ${\bf R}^4$.

Many of the space-time types of General Relativity are locally conformally flat and hence are M\"{o}bius structures, notably the de Sitter and anti de Sitter space-times. 

It is known that many Riemann surfaces can be viewed as a quotient of a discrete subgroup of $SL(2,{\bf R})$ acting on the complex upper half plane. Since $SL(2,{\bf R})\subset SL(2,{\bf C})\subset SU(2,2)$ the space of such Riemann surfaces can be imbedded in the category of M\"{o}bius structures. There is a large literature on conformal field theory in 2 dimensions in the context of statistical field theory and string theory.

The category of M\"{obius} structures has at least the richness of the category of 3-manifolds since, through the Thurston ``geometrization program" [5, 6] most 3-manifolds can be represented as a prime decomposition of model geometries where a model geometry is defined by the action of a discrete subgroup of a Lie group acting on a manifold. Lie groups involved in the model geometries are subgroups of $SU(2,2)$. Therefore the category of 3-manifolds can essentially be imbedded in the category of M\"{o}bius structures.

For the rest of this paper we will assume that the M\"{o}bius structures considered are orientable and orthochronous i.e. $J$ in Eq. \ref{eq:conformal_transformation} is the group generated by the orthochronous special Lorentz transformations and the dilatations.

The results described in this paper can be readily generalized from ELMCSs to LMCSs but are presented in the former context because of its familiarity.

Our work is related to the work of Dirac [7] who considered the derivation of physics in the context of 4 dimensional hypersurfaces in 5D projective space, generalizing the de Sitter universe. Our work is more general than that of Dirac. It is also related to the work of Cartan [8] who considered natural connections associated with homogeneous spaces. Our work is distinct to that of Cartan because, while some of the spaces studied by Cartan are M\"{o}bius structures there exist many M\"{o}bius structures which are not of this form.

Brozos-V\'{a}zquez {\em et al.} [9] point out that locally conformally flat space-times have not been studied much except the Schwarzschild interior solution and the Einstein static universe and also that Robertson-Walker space-times are conformally flat for any possible warping function. They obtain necessary and sufficient conditions for a static space-time to be locally conformally flat. Cabral and Lobo [10] discuss the connection between electrodynamics and the geometry and causal structure of space-time.  Schwarz  [11]  discusses the AdS/CFT correspondence and generalizations to d+1 dimensions. 

In the Riemannian case (as opposed to the pseudo-Riemannian case) an important class of locally conformally flat Riemannian $n$-manifolds arises when the developing map (conformal immersion into $S^n$) is injective in which case the manifold is the quotient of an open subset of $S^n$ by a Kleinian group [12]. Such locally conformally flat manifolds are called Kleinian manifolds and have been extensively studied by many mathematicians including Mostow, Thurston, Kulkarni, Goldman, Kamishima, Nayatani and Maier. Izeki [13] studied the interplay between the group cohomology of a Kleinian group and the topology of the associated locally conformally flat Riemannian manifold.

Locally conformally flat pseudo-Riemannian manifolds can be obtained in a similar fashion by taking open subsets of quotients of spaces such as $U(2)$ on which $SU(2,2)$ acts by M\"{o}bius automorphisms, by discrete subgroups of $SU(2,2)$.

Conformal transformations have been utilised in geometry and physics for more than a century [14].
The twistor program of Penrose [15] views space-time points as collections of null geodesics and obtains results concerning zero rest mass particles and fields.
't Hooft [16] considers a class of elementary particle models involving conformally flat space-times from the point of view of the conventional quantum gravity approach. 

``It may be necessary, for example, to reformulate classical geometry in a way that makes causal relations more fundamental, or to somehow `quantize' spacetime" [17]. Bombelli {\em et al.} [18] defined `causal sets' as locally finite partially ordered sets and proposed that space-time could be modeled as a causal set, and, in particular, as a discrete structure. This has led to a considerable amount of research appearing in the literature (the causal set program).

In this paper some properties of M\"{o}bius structures are described. As mentioned above M\"{o}bius structures are closely related to locally conformally flat (psuedo-) Riemannian manifolds. Five principal bundle structures may be naturally associated with a M\"{o}bius structure. The first has a structure group isomorphic to $O(1,3)^{\uparrow+}\times(0,\infty)$, the second has as structure group the conformal group. We consider a general class of principal bundles associated with M\"{o}bius structures where the transition functions are determined by $U(2,2)$ valued transition functions. Reduction in structure group from $G=U(2,2)$ to a subgroup $H$, say, is carried out. A natural epimorphism from $H$ to a subgroup $K\subset H\subset G$  results in a principal bundle $Q$, say.

Matter fields (fermion fields) are defined in the context of M\"{o}bius structures to be sections of vector bundles associated to $Q$. The case of the fundamental representation of the structure group $K$ of $Q$ is considered. Interaction fields are defined to be differential endomorphisms of the space of matter fields. Fundamental to this analysis is the use of an intertwining operator between a representation of $K$ on ${\bf R}^4$ and its adjoint representation on $u(2,2)$. It is shown that the Feynman slash operator (in the chiral representation of the Dirac gamma matrices) has just this intertwining property. It is shown how interaction fields can be generated by potential fields. Dirac's equation is shown to be an eigenvector equation associated with an interaction field endomorphism. The space of relative fields (relative to a reference field and an object which we call a gravitational gauge) is shown to have the structure of a bundle of operator algebras.

Maxwell's equations are derived by considering the canonical differential forms representing alternating multilinear forms which are obtained from the standard representation by the Hodge star operator. Distinguished alternating multilinear forms, representing physical electromagnetic fields, are obtained by considering the de Rham cohomology of forms.

\section{The conformal group}

In this section we summarize some properties of the conformal group [19], p. 64.

Let $C(1,3)$ denote the set of maximal domain conformal transformations in Minkowski
space. By Liouville's theorem $C(1,3)$ is (up to removable discontinuities) the 
set
of transformations generated by the translations, the Lorentz transformations,
the conformal inversion and the positive dilatations. $C(1,3)$ can be made into a
group by defining, for conformal transformations $h_1$, $h_2$, the product $h_1h_2$ to be
the unique $h \in C(1,3)$ such that $h \supset h_1 \circ h_2$, (where a function is identified with its graph).

Minkowski space can be identified with $u(2)$ by means of the Pauli algebra 
as follows.
\begin{equation}
\label{Pauli}
x \mapsto M(x) = ix^{\mu}\sigma_{\mu},
\end{equation}
where $\{\sigma_{\mu}\}_{\mu = 1}^3$ are the Pauli matrices
\[ \sigma_1 = \left(\begin{array}{cc} 0 & 1 \\ 1 & 0 \end{array} \right),
   \sigma_2 = \left(\begin{array}{cc} 0 & -i \\ i & 0 \end{array} \right),
   \sigma_3 = \left(\begin{array}{cc} 1 & 0 \\ 0 & -1 \end{array} \right), \]
and $\sigma^0 = 1_2$. The map $M$ has the property that 
\begin{equation}
Q(x) = -\det(M(x)), \forall x\in {\bf R}^4,
\end{equation}
where $Q$ is the Minkowski space quadratic form.

$U(2,2)$ is defined to be the set of all matrices $A\in{\bf C}^{4\times 4}$ such that
\begin{equation}
A^{\dagger}gA = g,
\end{equation}
where $g$ is a Hermitian form of signature $(2,2)$ and $SU(2,2)$ is defined by
\begin{equation}
SU(2,2)=\{A\in U(2,2):\mbox{det}(A)=1\}.
\end{equation}
If $U(2,2)$ is considered in the representation in which the 
Hermitian form $g$ is defined by the matrix
\begin{equation}
g=\left(
\begin{array}{cc}
    0 & 1_2 \\
    1_2 & 0 
\end{array} 
\right),\label{eq:Minkowski}
\end{equation}
then, for each $A \in U(2,2)$ the map $f_{1,A}$ defined in Minkowski ($u(2)$) space by 
\begin{equation}
f_{1,A}(M) = (aM+b)(cM+d)^{-1}, \label{eq:transformation} 
\end{equation}
where
\[ A = \left(
\begin{array}{llll}
    a & b \\
    c & d 
\end{array}
\right), \]
is an element of $C(1,3)$. Furthermore, the map $f_1 = (A\mapsto f_{1,A})$ is a homomorphism from $U(2,2)$ to $C(1,3)$. The map $f_1|_{SU(2,2)}$ has kernel $\{\pm 1, \pm i\}$ and its image contains the proper orthochronous conformal group $C(1,3)^{\uparrow +}$.

The map $A\mapsto f_{1,A}$ has the property that
\begin{equation}
f_{1,\mu A}=f_{1,A}, \forall \mu\in U(1), A\in U(2,2).
\end{equation} 

\section{The principal bundle structures for a M\"{o}bius structure}

In this section we describe five principal bundles which may be associated with a M\"{o}bius structure.  The first has as structure group the Lorentz group times the positive dilatations. The second has as structure group the conformal group. We assume that the transition functions for this bundle can be obtained as images of the transition functions of a bundle with structure group $U(2,2)$. The fourth is obtained from the third by reduction of structure group and the fifth is a homomorphic image of the fourth.

The first way that a  M\"{o}bius structure can be considered as a principal bundle is as follows.
Let $d_{ij}:U_{i}\cap U_{j}\rightarrow J$, where
\begin{equation}
J = \{ \lambda\Lambda : \lambda\in(0,\infty),\Lambda\in O(1,3)^{\uparrow+} \},
\end{equation}
be defined by,
\begin{equation} \label{eq:transformation_derivative}
d_{ij}(x) = (D(\phi_{i}\circ\phi_{j}^{-1}))(\phi_{j}(x)).
\end{equation}
Let $R = (R,X,J,\pi)$ be the principal bundle obtained by taking the $\{ d_{ij} \}$ as transition functions.
The structure group of this bundle is $J \cong O(1,3)^{\uparrow+}\times (0,\infty)$. The vector bundle associated to the fundamental representation of $J$ is isomorphic to the tangent bundle of $X$.
Define functions $\lambda_{ij}:U_{i}\cap U_{j}\rightarrow (0,\infty)$ and $\Lambda_{ij}:U_{i}\cap U_{j} \rightarrow O(1,3)^{\uparrow+}$ by 
\begin{equation}
\label{lambda1}
\lambda_{ij}(x) = (\det(d_{ij}(x)))^{\frac{1}{4}},
\end{equation} 
\begin{equation} \label{eq:Lorentz_scale}
\Lambda_{ij}(x) = \lambda_{ij}(x)^{-1}d_{ij}(x).
\end{equation}

Given a conformal transformation $f$ defined in Minkowski space let $C(f)$
denote the unique element $h \in C(1,3)$ such that $f \subset h$. $f \mapsto
C(f)$ has the property that 
\begin{equation}
\forall g_1,g_2\in\Gamma(1,3),g_1 \circ g_2 \neq \emptyset \Rightarrow C(g_1 \circ g_2) = C(g_1)C(g_2). 
\end{equation}
Let $(X,{\cal A})$ be a M\"{o}bius structure. Suppose that for each $i,j 
\in I,\mu_{ij}$ is defined by 
\begin{equation}
\mu_{ij} = C(\phi_i \circ \phi_j^{-1}).
\end{equation}
If we write $\mu_{ij}(x) = \mu_{ij}$ then $\mu_{ij}$ can be thought of as a
(constant) $C(1,3)^{\uparrow +}$ valued function over $U_i \cap U_j$. It is straightforward to show that the cocycle condition
\begin{equation}
\mu_{ij}(x) \mu_{jk}(x) = \mu_{ik}(x),
\end{equation}
is satisfied.
Therefore $\{\mu_{ij}\}$ define transition functions for a principal bundle \newline
$B = (B,X,C(1,3)^{\uparrow+},\pi)$ with typical fiber $C(1,3)^{\uparrow+}$. 

$C(1,3)^{\uparrow+}$ is locally isomorphic to $SU(2,2)$. For the rest of this paper we will investigate the consequences of the assumption that there are $U(2,2)$ valued (constant) transition functions 
$\{g_{ij}\} \subset U(2,2)$ such that for all $i,j\in I$ such that $U_i\cap U_j\neq\emptyset$ 
\begin{equation}
\mu_{ij}=C(\phi_i\circ\phi_j^{-1}) = f_{1,g_{ij}}.
\end{equation}
This assumption is analogous to the assumption of orientability that we have made above which as frequently made in differential geometric investigations. Thus we are considering a subclass of the class of M\"{o}bius structures. 
A consequence of this assumption is that the Lie groups that will be considered in this work
are matrix Lie groups. 

Let $G = U(2,2)$ and $P_G = (P_G,X,G,\pi)$ be the principal fiber bundle obtained by taking $\{g_{ij}\}$ as transition functions.
 
One would like to determine a natural principal bundle structure for a given
M\"{o}bius
structure. An indication
of what this principal bundle structure might be can be obtained by
considering the M\"{o}bius structure $U(2)$.
In the representation for
$U(2,2)$ in which the Hermitian form is defined by the matrix
\begin{equation}
g = \left(
\begin{array}{cc}
    1 & 0 \\
    0 & -1 
\end{array}
\right), \label{eq:U(2)}
\end{equation}
$U(2,2)$ acts on $U(2)$ by
\begin{equation}
f_{2,A}(u) = (au+b)(cu+d)^{-1}. \label{eq:action}
\end{equation}
Consider the Cayley transform $C : u(2) \rightarrow U(2)$ defined by
\begin{equation}
C(M) = (1-M)(1+M)^{-1},
\end{equation}
and its inverse, with domain an open subset of $U(2)$, defined by
\begin{equation}
C^{-1}(u) = (1-u)(1+u)^{-1}.
\end{equation}
If $U(2)$ is given the atlas consisting of the inverse Cayley transform and its 
images as a result of being acted on by elements of $U(2,2)$ (where, as usual, Minkowski space ${\bf R}^4$ has been identified with $u(2)$) then $U(2,2)$ acts by M\"{o}bius automorphisms. The
isotropy subgroup at $e$ is the group
\begin{equation}
H = \{ \left(
\begin{array}{cc}
    a & b \\
    c & d 
\end{array}
\right) \in U(2,2) : a+b = c+d \}.
\end{equation}
Therefore $U(2) \approx U(2,2)/H$. It follows that $U(2,2)$ can be viewed as 
the total space of a principal fiber bundle with base space $U(2)$ and typical fiber $H$.

With respect to the representation for $U(2,2)$ where the metric $g$ is defined by Equation~\ref{eq:Minkowski} $H$ is given by
\begin{equation}
H = \{ \left(
\begin{array}{ll}
a & b \\
c & d
\end{array} \right) \in U(2,2) : b = 0 \},
\end{equation}
and 
\begin{equation}
\forall A\in U(2,2), ((0\in \mbox{Domain}(f_{1,A}) \mbox{ and } f_{1,A}(0) = 0) \Leftrightarrow A \in H).
\end{equation}

\subsection{Reduction of structure group from $G$ to $H$}

We take the point of view
that geometric objects at a point in space-time are things which
have values in all coordinate systems about that point and which transform 
covariantly. In special relativistic physics ``covariantly'' is taken to be with respect to the group $O(1,3)$ of Lorentz transformations while in general relativity one considers ``general covariance" which is covariance with respect to the group of (germs of) diffeomorphisms between neighbourhoods of the point in space-time under consideration. In our work we restrict the diffeomorphisms to be conformal transformations. 

If $X$ is a smooth manifold with atlas ${\mathcal A}=\{(U_i,\phi_i):i\in I\}$ and $P_1 = (P_1,X,G_1,\pi)$ is a principal bundle with structure group $G_1$ and $\rho : G_1 \times V \rightarrow V$ is any representation of $G_1$
as automorphisms of an object $V$ then there is an associated bundle $E=\bigcup_{x\in X}E_x$ with typical fiber $V$ and elements $v\in E_x$ can be thought of as maps $v : I_{x}
\rightarrow V$, where $I_{x} = \{ i\in I : x \in U_{i} \}$,
such that
\[ v_{i} = \rho(g_{1ij}(x)) v_{j}, \forall i,j\in I_x,\]
where $\{g_{1ij}\}$ are the transition functions for $ P_1$  and $v_{i}$ denotes $v(i)$ the value of $v$ in coordinate system $i$.

\newtheorem{theorem}{Theorem}
\begin{theorem} \label{reduction1}
Let $X$ be a $C^{\infty}$  manifold with atlas ${\mathcal A}=\{(U_i,\phi_i):i\in I\}$, $G_1$ be a Lie group,  $\{g_{1ij}\}$ be $G_1$ valued transition functions for a principal bundle $R_1 = (R_1,X,G_1,\pi)$ on $X$ and $H_1$ be a Lie subgroup of $G_1$. Then if $g_i : U_i \rightarrow G_1$ are smooth for all $i$ and are such that 
\begin{equation} \label{eq:Theorem1}
h_{ij}(x) = g_i(x)^{-1}g_{1ij}(x)g_j(x) \in H_1, \forall x\in U_i\cap U_j, \forall i,j\in I,
\end{equation}
then $\{ h_{ij} \}$ form transition functions for a principal bundle with structure group $H_1$ which can be obtained from $R_1$ by reduction of structure group. Conversely if $R_2 = (R_2,X,H_1,\pi_1)$ is a principal bundle and $f:R_2\rightarrow R_1$ is a reduction of structure group from $R_1$ to $R_2$ then there exist smooth $g_i:U_i \rightarrow G_1$ such that Equation~\ref{eq:Theorem1} holds and ${h_{ij}}$ are the transition functions for $R_2$.
\end{theorem}
A proof of this theorem can be found in [2]. Let 
\[ C_0(1,3) = \{ f\in C(1,3) : 0\in \mbox{Domain}(f), f(0) = 0 \}. \]
\begin{theorem} \label{reduction}
Let $X$ be a M\"obius structure. Then there is a reduction of structure group
$\alpha:B_0 \rightarrow B$ from $C(1,3)$ to $C_0(1,3)$ where $B_0$ is a principal fiber
bundle with structure group $C_0(1,3)$. Also there exists a reduction of structure group $\beta:P\rightarrow P_G$ from $G$ to $H$ where $P$ is a principal bundle with structure group $H$. The transition functions $\{ \zeta_{ij} \}$ for $B_0$ are related to the transition functions $\{ h_{ij} \}$ for $P$ by
\[ \zeta_{ij} = f_{1,h_{ij}}. \] 
\end{theorem}
{\bf Proof}
Let $\tau_a$ for $a \in {\bf R}^4$ be the translation operator defined by
\[ \tau_a(b) = b - a. \]
Define
\[ \zeta_{ij}(x) = C((\tau_{\phi_{i}(x)} \circ \phi_i) \circ (\tau_{\phi_{j}(x)}
\circ \phi_j)^{-1}). \]
Each $\zeta_{ij}(x)$ is an element of $C(1,3)$ such that $(\zeta_{ij}(x))(0) = 0$. 
Therefore $\zeta_{ij}(x)
\in C_0(1,3), \forall i,j$ and $x \in U_i \cap U_j$. Now
\begin{equation}
\zeta_{ij}(x) = C(\tau_{\phi_i(x)})C(\phi_i\circ\phi_j^{-1})C(\tau_{\phi_j(x)}^{-1}).
\end{equation}
Therefore by Theorem~\ref{reduction1} $\{\zeta_{ij}\}$ define
transition functions for a principal bundle $B_0$ which can be obtained from $B$ by reduction of structure group from $C(1,3)$ to $C_0(1,3)$.

Furthermore
\begin{eqnarray}
\zeta_{ij}(x) & = & C(\tau_{\phi_i(x)})C(\phi_i\circ\phi_j^{-1})C(\tau_{\phi_j(x)}^{-1}) \nonumber \\
    & = & \tau_{\phi_i(x)}C(\phi_i\circ\phi_j^{-1})\tau_{\phi_j(x)}^{-1} \nonumber \\
    & = & f_{1,g_i(x)}\circ f_{1,g_{ij}}\circ f_{1,g_j^{-1}(x)} \nonumber \\
    & = & f_{1,g_i(x)g_{ij}g_j^{-1}(x)}, \nonumber
\end{eqnarray}
where
\[ g_i(x) = \left(\begin{array}{cc} 1 & -\phi_i(x) \\ 0 & 1 \end{array}\right)\in U(2,2), \]
where, as usual, the point $\phi_i(x)\in V_i$ has been identified with its corresponding point in $u(2)$. Therefore
\[ \zeta_{ij}(x) = f_{1,h_{ij}(x)}, \]
where
\[ h_{ij}(x) = g_i(x)g_{ij}g_j^{-1}(x). \]
Now (viewing an element $A\in U(2,2)$ as being the map $f_{1,A}$ defined in a subset of $u(2)$) we have $g_i(x)(M)=M-\phi_i(x),\forall x\in U_i,M\in u(2),i\in I$.  Thus $g_i(x)(\phi_i(x))=0$ and so $(g_i(x))^{-1}(0)=\phi_i(x), \forall x\in U_i,i\in I$. Thus 
\[h_{ij}(x)(0)=(g_i(x)C(\phi_i\circ\phi_j^{-1}))(\phi_j(x))=g_i(x)(\phi_i(x))=0, \forall x\in U_i\cap U_j, i,j\in I. \] 
Hence $h_{ij}(x)\in H, \forall x\in X, i,j\in I$. 

Therefore by Theorem~\ref{reduction1} $\{ h_{ij} \}$ form transition functions for a principal bundle $P$ which can be obtained from $P_G$ by a reduction of structure group from $G$ to $H$. $\Box$

This reduction of structure group from $G$ to $H$ will be called the standard 
reduction from $G$ to $H$. 

In a subsequent paper it will be shown that there is a map from $T_x P$ to ${\mathfrak g}=u(2,2)$ for all $x\in X$ which transforms under the adjoint action of ${\mathfrak g}$ under a change of coordinate system. Therefore $TP$ has the structure of a bundle of Lie algebras and can be made into a psuedo-Riemannian manifold when equipped with the metric induced by the Killing form in ${\mathfrak g}$. 

\subsection{The principal bundle $Q$ resulting from a natural epimorphism from $H$ to a subgroup $K\subset H$}

Let
\begin{equation}
K  =  
  \{ \left(\begin{array}{cc}
a & 0 \\
0 & a^{\dagger -1}
\end{array}\right) : a\in GL(2,{\bf C}), |\mbox{det}(a)|=1\}. \nonumber
\end{equation}
It is straighfoward to show that $K\subset H$.
$K$ is isomorphic to $SL(2,{\bf C})\times U(1)$. 

If $\kappa\in K$ then $f_{1,\kappa}$ is a Lorentz transformation. This is because $f_{1,\kappa}$ is a linear map from $u(2)$ to $u(2)$ and
\begin{eqnarray}
Q(f_{1,\kappa}(M)) & = & -\det(f_{1,\kappa}(M)) \nonumber \\
 & = & -\det(aMa^{\dagger}) \nonumber \\
 & = & -|\det(a)|^2\det(M) \nonumber \\
 & = & Q(M), \nonumber
\end{eqnarray}
for all $M\in u(2)$ where
\[ \kappa = \left(\begin{array}{cc} a & 0 \\ 0 & a^{\dagger -1} \end{array}\right). \]
 Now $A\mapsto f_{1,A}$ is a homomorphism from $U(2,2)$ to $C(1,3)$. Thus the map $\kappa\mapsto f_{1,\kappa}$ is a homomorphism and defines an action of $K$ on Minkowski space. We will call the representation of $K$ on Minkowski space induced by this map the standard representation of $K$ on Minkowski space. If $\kappa\in K$ then denote by $\Lambda(\kappa)$ the Lorentz transformation associated with $\kappa$.
 
If $\left(\begin{array}{cc}
a & 0 \\
c & d
\end{array}\right) \in H$ then
\[ \left(\begin{array}{cc}
a^{\dagger} & c^{\dagger} \\
0 & d^{\dagger}
\end{array}\right)
\left(\begin{array}{cc}
0 & 1 \\
1 & 0
\end{array}\right)
\left(\begin{array}{cc}
a & 0 \\
c & d
\end{array}\right) = 
\left(\begin{array}{cc}
0 & 1 \\
1 & 0
\end{array}\right). \]
Therefore
\begin{equation}
a^{\dagger}d = 1\mbox{ and } c^{\dagger}a+a^{\dagger}c = 0. \label{H_condition}
\end{equation}
Thus
\[ d = a^{\dagger -1}, \]
and
\[ \mbox{det}(a) \neq 0. \]
Hence we may define $\Theta : H \rightarrow K$ by
\begin{equation}
\Theta(\left(\begin{array}{cc}
a & 0 \\
c & d
\end{array} \right)) =
\left( \begin{array}{cc}
\lambda^{-\frac{1}{2}}a & 0 \\
0 & \lambda^{\frac{1}{2}} d
\end{array}\right), \lambda = |\det(a)|.
\end{equation}
It is straightforward to show that $\Theta$ is a well defined homomorphism. For each $i,j\in I$ and $x \in U_i\cap U_j$ define
\begin{equation}
\kappa_{ij}(x) = \Theta(h_{ij}(x)).
\end{equation}
Then $\{ \kappa_{ij} \}$ form transition functions for a principal bundle 
$Q = (Q,X,K,\pi)$ with structure group $K$.

\begin{theorem}
\label{reduction3}
If $\kappa_{ij}$ have been obtained as described above using the standard reduction $f : P \rightarrow P_G$
then the induced transformation $f_{1,\kappa_{ij}(x)}$ of Minkowski space (identified with $\mbox{u}(2)$) is $\Lambda_{ij}(x)$.
\end{theorem}
{\bf Proof}
Let $ A=\left(\begin{array}{cc} a & b \\ c & d \end{array}\right) = g_{ij}(x) = g_{ij}$. Then
\begin{eqnarray}
h_{ij}(x)  
  & = & \left(\begin{array}{cc} 1 & -\phi_i(x) \\ 0 & 1 \end{array}\right)
        \left(\begin{array}{cc} a & b \\ c & d \end{array}\right)
        \left(\begin{array}{cc} 1 & \phi_j(x) \\ 0 & 1 \end{array}\right) \nonumber \\
  & = & \left(\begin{array}{cc} a-\phi_i(x)c & (a-\phi_i(x)c)\phi_j(x)+b-\phi_i(x)d \\ c & c\phi_j(x)+d \end{array}\right) \nonumber
\end{eqnarray}
Now
\[ (a-\phi_i(x)c)\phi_j(x)+b-\phi_i(x)d = -\phi_i(x)(c\phi_j(x)+d)+a\phi_j(x)+b = 0, \]
because
\[ \phi_i(x) = f_{1,A}(\phi_j(x)) = (a\phi_j(x)+b)(c\phi_j(x)+d)^{-1}. \]
Therefore
\begin{equation}
\kappa_{ij}(x) = \left(\begin{array}{cc} \lambda^{-\frac{1}{2}}( a-\phi_i(x)c) & 0 \\ 
0 & \lambda^{\frac{1}{2}}(c\phi_j(x)+d) \end{array} \right),
\end{equation}
where
\[ \lambda = |\det(a-\phi_i(x)c)|. \]
Therefore
\begin{equation}
f_{1,\kappa_{ij}(x)}(N) = \lambda^{-1}(a-\phi_i(x)c)N(c\phi_j(x)+d)^{-1}.
\end{equation}
It can be shown that the map $f_{1,A}$ has derivative
\[ (f_{1,A}^{\prime}(M))(N) = (a-f_{1,A}(M)c)N(cM+d)^{-1}. \]
Therefore $\Lambda_{ij}(x)=f_{1,\kappa_{ij}(x)}$  and, comparing with Eq. \ref{eq:Lorentz_scale}, $\lambda_{ij}(x)=\lambda$. $\Box$

\subsection{Some other principal bundles associated with a M\"{o}bius structure $X$}

By Kobayashi and Nomizu [3] p. 59 the structure group $H$ of $P$ is reducible to any of its maximal compact subgroups. Such maximally compact subgroups are locally isomorphic to $U(1)\times SU(2)\times U(1)$.

\section{Fermion fields and interaction fields}

We have discussed M\"{o}bius structures and some of their properties. In particular, there is, with any M\"{o}bius structure $(X,{\mathcal A})$ associated, in a natural way, a principal bundle $Q$. $Q$ has structure group $K$ which is isomorphic to $SL(2,{\bf C})\times U(1)$. 

One may describe a general paradigm for generating physical theories as follows. Take any space $V$ (e.g. a vector space) and and a representation of $K$ in the space of endomorphisms (linear maps) in $V.$ Construct the (vector) bundle $E$ associated with the representation and consider the space Sec$(E)$ of sections of the bundle. 

Physical fields and particles (fermions) are identified with elements of Sec$(E)$. We propose that interaction fields (bosons) can be  identified with endomorphisms of Sec$(E),$ i.e. mappings from Sec$(E)$ to Sec$(E)$ which are linear.

Our approach does not derive physics using the commonly used variational principles (involving Hamiltonians, Lagrangians, gravitational actions etc.) but focuses on mathematically ``natural", and in particular, well defined, constructions and could be readily axiomatized from the starting point of LMCSs with few axioms.

Nevertheless, the action principles remain true since while the action principles imply the field equations, the field equations imply the action principles. We just do not take the action principles as the starting point from which physics is derived.

There is a large literature on the action of conformal vector fields on spinor bundles (e.g. [20, 21]) and conformally invariant differential operators on Minkowski space [22]. Our work is different to this work, we do not construct differential operators with respect to given vector fields on the manifold (Lie derivatives). Rather, we start with an intertwining operator between the standard representation of $K$ on ${\bf R}^4$ and its adjoint representation in $u(2,2)$ and construct endomorphisms of Sec$(E)$. We use the fact, here and in all the following work, that $K$ acts on ${\bf R}^4$, ${\bf C}^4$ and $u(2,2)$ in natural ways. We show that, remarkably, the Feynman slash operator (in its chiral representation) has just the intertwining properties that we require (and, in particular, the Dirac gamma matrices, when multiplied by the imaginary unit $i$, are all elements of $u(2,2)).$ Dirac's equation is shown to be the eigenvector equation for the endomorphisms that we have constructed.

The space of interaction fields is shown to have an affine structure, given a reference field and an object which we call a gravitational gauge, there is a natural mapping between the space of interaction fields and the space of sections of a bundle of operator algebras.

 \section{Analytic properties of fermion fields and interaction fields on M\"{o}bius structures}

In this section we define matter fields and interaction fields in the context of the associated vector bundle $E$ to the principle bundle $Q$ through the fundamental representation of $K$. Central to this section are operators $\Sigma^{\mu}$ which will be later related to the Dirac gamma matrices in the chiral representation but which are here defined by their property as  intertwining operators. The properties of these intertwining operators together with the properties of the M\"{o}bius structure coordinate transformations lead to simple transformation properties of differential operators constructed using them. Interaction fields allow for the construction of differential operators which have natural transformation properties and which are, up to multiplication by a (gravitational) gauge, differential endomorphisms of the space of matter fields.

Let Sec$(E)$ denote the space of C$^{\infty}$ sections 
of $E$. Elements of Sec$(E)$ will be called fermion fields, matter fields or simply fields.
Elements of Sec$(E)$ can be thought of as collections $\psi=\{\psi_i\}_{i\in I}$ which satisfy
\[ \psi_{i} \in \mbox{C}^{\infty}(V_i,{\bf C}^{4}), \forall i \in I, \mbox{and}\]
\begin{equation}
 \psi_{i} = (\kappa_{ij}\circ\phi_{i}^{-1})(\psi_{j}\circ (\phi_{j}\circ\phi_{i}^{-1})), \forall i,j\in I \mbox{ for which }U_i\cap U_j\neq\emptyset.
\label{eq:matter_field}
\end{equation}
Sec$(E)$ is a module over the algebra C$^{\infty}(X,{\bf C})$.

Let ${\mathfrak g}=u(2,2)$ be the Lie algebra of $U(2,2)$.
Suppose that $\Sigma : {\bf R}^{4} \rightarrow {\mathfrak g}$ is an intertwining operator between the standard representation of $K$ in ${\bf R}^{4}$ and the adjoint representation of $K$ in $\mathfrak g$. An explicit example of such an intertwining operator will be given in a later section.
 Then for all $\kappa \in K$
\begin{equation} \label{eq:intertwine}
\kappa\Sigma(u)\kappa^{-1} = \Sigma(\Lambda u), \forall u \in
{\bf R}^{4},
\end{equation}
where $\Lambda=f_{1,\kappa}$ is the Lorentz transformation of Minkowski space
corresponding to $\kappa$. Therefore
\[ \kappa\Sigma(u^{\mu}e_{\mu}) = \Sigma({\Lambda^{\mu}}_{\nu}u^{\nu}e_{\mu})\kappa. \]
From this it follows that
\[ \kappa\Sigma_{\mu} = {\Lambda^{\nu}}_{\mu}\Sigma_{\nu}\kappa, \]
where $\Sigma_{\mu} = \Sigma(e_{\mu})$ for $\mu = 0,1,2 ,3$ and $\{ e_{\mu} \}_{\mu = 0}^{3}$ is the standard basis for ${\bf R}^{4}$. Let $\Sigma^{\mu}$ be obtained from $\Sigma_{\nu}$ by raising of indices. That is
\[ \Sigma^{\mu} = \eta^{\mu\nu}\Sigma_{\nu}, \]
where $\eta$ is the Minkowski space metric. Then
\[ \kappa\eta_{\mu\alpha}\Sigma^{\alpha} = {\Lambda^{\nu}}_{\mu}\eta_{\nu\beta}\Sigma^{\beta}\kappa. \]
Hence
\begin{eqnarray}
\kappa\Sigma^{\rho} & = & \eta^{\rho\mu}{\Lambda^{\nu}}_{\mu}\eta_{\nu\beta}\Sigma^{\beta}\kappa \nonumber \\
  & = & \eta^{\rho\mu}{{\Lambda^T}_{\mu}}^{\nu}\eta_{\nu\beta}\Sigma^{\beta}\kappa \nonumber \\
  & = & {{\Lambda^{-1}}^{\rho}}_{\beta}\Sigma^{\beta}\kappa. \nonumber
\end{eqnarray}
Therefore
\begin{equation} \label{eq:intertwine1}
\kappa\Sigma^{\nu} = {{\Lambda^{-1}}^{\nu}}_{\mu}\Sigma^{\mu}\kappa. 
\end{equation}
\begin{theorem} \label{proposition:intertwine1}
Matter fields $\psi \in \mbox{Sec}(E)$ have the following differential transformation property.
\begin{eqnarray}
\Sigma^{\mu}\partial_{\mu}\psi_{i} & = &
(\lambda_{ij}^{-1}\circ\phi_{i}^{-1})(\kappa_{ij}\circ\phi_{i}^{-1})\Sigma^{\nu}((\partial_{\nu}\psi_{j})\circ (\phi_{j}\circ\phi_{i}^{-1})) +
\Sigma^{\mu}\partial_{\mu}(\kappa_{ij}\circ\phi_{i}^{-1}) \nonumber \\ 
    & & \mbox{    } (\psi_{j}\circ(\phi_{j}\circ\phi_{i}^{-1})). \nonumber
\end{eqnarray}
\end{theorem}
{\bf Proof}
\begin{eqnarray}
\Sigma^{\mu}\partial_{\mu}\psi_{i} & = &
 \Sigma^{\mu}\partial_{\mu}((\kappa_{ij}\circ\phi_{i}^{-1})(\psi_{j}
\circ(\phi_j\circ \phi_{i}^{-1}))) \nonumber \\
& = & \Sigma^{\mu}(\kappa_{ij}\circ\phi_{i}^{-1})\partial_{\mu}(\psi_{j}\circ(\phi_j\circ
    \phi_{i}^{-1})) +
 \Sigma^{\mu}\partial_{\mu}(\kappa_{ij}\circ\phi_{i}^{-1})(\psi_{j}\circ(\phi_{j}\circ\phi_{i}^{-1})) \nonumber \\
    & = & \Sigma^{\mu}(\kappa_{ij}\circ\phi_{i}^{-1})(\lambda_{ji}\circ\phi_{i}^{-1}){({\Lambda_{ji}\circ\phi_{i}^{-1})}^{\nu}}_{\mu}((\partial_{\nu}\psi_{j})
    \circ (\phi_{j}\circ\phi_{i}^{-1})) + \nonumber \\
    & & \Sigma^{\mu}\partial_{\mu}(\kappa_{ij}\circ\phi_{i}^{-1})(\psi_{j}\circ(\phi_{j}\circ\phi_{i}^{-1})) \nonumber \\
    & = & (\lambda_{ij}^{-1}\circ\phi_{i}^{-1})\Sigma^{\mu}(\kappa_{ij}\circ\phi_{i}^{-1}){({\Lambda_{ij}^{-1}\circ\phi_{i}^{-1})}^{\nu}}_{\mu}((\partial_{\nu}\psi_{j})
    \circ (\phi_{j}\circ\phi_{i}^{-1})) + \nonumber \\
    & & \Sigma^{\mu}\partial_{\mu}(\kappa_{ij}\circ\phi_{i}^{-1})(\psi_{j}\circ(\phi_{j}\circ\phi_{i}^{-1})) \nonumber \\
    & = & (\lambda_{ij}^{-1}\circ\phi_{i}^{-1})(\kappa_{ij}\circ\phi_{i}^{-1})\Sigma^{\nu}((\partial_{\nu}\psi_{j})\circ (\phi_{j}\circ\phi_{i}^{-1})) +
\Sigma^{\mu}\partial_{\mu}(\kappa_{ij}\circ\phi_{i}^{-1}) \nonumber \\
    & & (\psi_{j}\circ(\phi_{j}\circ\phi_{i}^{-1})), \nonumber 
\end{eqnarray}
\mbox{  } $\Box$ \newline
where the last equality occurs by virtue of the identity Equation~\ref{eq:intertwine1} which results in the disappearance of $\Lambda$ from the expression for $\Sigma^{\mu}\partial_{\mu}\psi_i$.

Define an interaction field to be a collection $\Phi = \{ \Phi_{i} \}_{i \in I}$ of quantities \newline
$\Phi_{i}\in C^{\infty}(V_i,\mbox{End}({\bf C}^{4}))$ which transform according to
\begin{eqnarray}
\Phi_{i} & = & (\lambda_{ij}^{-1}\circ\phi_{i}^{-1})(\kappa_{ij}\circ\phi_{i}^{-1})(\Phi_{j}\circ (\phi_j\circ\phi_{i}^{-1}))(\kappa_{ij}^{-1}\circ\phi_{i}^{-1}) 
\nonumber \\
    & & \mbox{        } + \Sigma^{\mu}\partial_{\mu}(\kappa_{ij}\circ\phi_{i}^{-1})(\kappa_{ij}^{-1}\circ\phi_{i}^{-1}).
\end{eqnarray}

Let ${\cal F}$ denote the set of all interaction fields. 

\begin{theorem} \label{Dirac2}
Suppose that $\Phi\in{\cal F}$. Then the map $T_{\Phi}$ defined on Sec$(E)$ 
by
\begin{equation}
(T_{\Phi}\psi)_{i} = (\Sigma^{\mu}\partial_{\mu} - \Phi_{i})\psi_{i},
\end{equation}
has the following transformation property.

\begin{equation}
(T_{\Phi}\psi)_{i} = (\lambda_{ij}^{-1}\circ\phi_{i}^{-1})(\kappa_{ij}\circ\phi_{i}^{-1})((T_{\Phi}\psi)_{j}\circ(\phi_{j}\circ\phi_{i}^{-1})).
\end{equation}

\end{theorem}

{\bf Proof}
\begin{eqnarray}
(T_{\Phi}\psi)_{i} & = & (\Sigma^{\mu}\partial_{\mu}-\Phi_{i})\psi_{i} 
    \nonumber \\
    & = & \Sigma^{\mu}\partial_{\mu}\psi_{i}-\Phi_{i}\psi_{i}
    \nonumber \\
    & = & (\lambda_{ij}^{-1}\circ\phi_{i}^{-1})(\kappa_{ij}\circ\phi_{i}^{-1})\Sigma^{\nu}((\partial_{\nu}\psi_{j})\circ(\phi_{j}\circ\phi_{i}^{-1}))
    +\Sigma^{\mu}\partial_{\mu}(\kappa_{ij}\circ\phi_{i}^{-1}) \nonumber \\
    & & \mbox{    } (\psi_{j}\circ(\phi_{j}\circ\phi_{i}^{-1}))
    -((\lambda_{ij}^{-1}\circ\phi_{i}^{-1}) (\kappa_{ij}\circ\phi_{i}^{-1})(\Phi_{j}\circ (\phi_{j}\circ\phi_{i}^{-1}))(\kappa_{ij}^{-1}\circ\phi_{i}^{-1})+
\nonumber \\
    & & \mbox{    } \Sigma^{\mu}\partial_{\mu}(\kappa_{ij}\circ\phi_{i}^{-1})(\kappa_{ij}^{-1}\circ\phi_{i}^{-1}))(\kappa_{ij}\circ\phi_{i}^{-1})(\psi_{j}\circ (\phi_{j}\circ\phi_{i}^{-1})) \nonumber \\
    & = & (\lambda_{ij}^{-1}\circ\phi_{i}^{-1})(\kappa_{ij}\circ\phi_{i}^{-1})(\Sigma^{\nu}((\partial_{\nu}\psi_{j})\circ(\phi_{j}\circ\phi_{i}^{-1}))
    - (\Phi_{j}\circ (\phi_{j}\circ\phi_{i}^{-1}))
\nonumber \\
    & & \mbox{    } (\psi_{j}\circ(\phi_{j}\circ\phi_{i}^{-1}))) \nonumber \\
    & = & (\lambda_{ij}^{-1}\circ\phi_{i}^{-1})(\kappa_{ij}\circ\phi_{i}^{-1})(((\Sigma^{\nu}\partial_{\nu}-\Phi_{j})\psi_{j})\circ (\phi_{j}\circ\phi_{i}^{-1})) \nonumber \\
    & = & (\lambda_{ij}^{-1}\circ\phi_{i}^{-1})(\kappa_{ij}\circ\phi_{i}^{-1})
((T_{\Phi}\psi)_{j}\circ (\phi_{j}\circ\phi_{i}^{-1})). \nonumber
\end{eqnarray}
\mbox{  } $\Box$ \newline

Define a {\em gauge} (gravitational gauge) to be a section of the principal $(0,\infty)$ bundle obtained by taking $\{ \lambda_{ij} \}$ as transition functions. Since $(0,\infty)$ is diffeomorphic to a Euclidean space it follows [3] that there exists at least one gauge. Let the set of gauges be denoted by ${\cal G}$. Gauges can be thought of as collections $\zeta = \{\zeta_i\}$ of functions $\zeta_i\in C^{\infty}(V_i,(0,\infty))$ which transform according to
\begin{equation} \label{eq:gauge}
\zeta_i(\xi) = (\lambda_{ij}\circ\phi_i^{-1})(\xi)(\zeta_j\circ\phi_j\circ\phi_i^{-1})(\xi).
\end{equation}
We have the following.
\begin{theorem} \label{Dirac3}
Let $\Phi\in {\cal F}$ and $\zeta \in {\cal G}$. Then the map $T_{\Phi,\zeta}$ defined on Sec$(E)$ by
\begin{equation}
(T_{\Phi,\zeta}\psi)_{i} = \zeta_{i}(T_{\Phi}\psi)_{i},
\end{equation}
is a linear endomorphism of Sec$(E)$.
\end{theorem}
{\bf Proof}
\begin{eqnarray}
(T_{\Phi,\zeta}\psi)_{i} & = & \zeta_{i}(T_{\Phi}\psi)_{i} \nonumber \\
    & = & (\lambda_{ij}\circ\phi_{i}^{-1})(\zeta_{j}\circ\phi_{j}\circ\phi_{i}^{-1})(\lambda_{ij}^{-1}\circ\phi_{i}^{-1})(\kappa_{ij}\circ\phi_{i}^{-1})((T_{\Phi}\psi)_{j}\circ(\phi_{j}\circ\phi_{i}^{-1})) \nonumber \\
    & = & (\kappa_{ij}\circ\phi_{i}^{-1})((T_{\Phi,\zeta}\psi)_{j}\circ\phi_{j}\circ\phi_{i}^{-1}). \nonumber 
\end{eqnarray}
Hence $T_{\Phi,\zeta}\psi$ is a matter field, by the defining property of Equation~\ref{eq:matter_field}. Therefore $T_{\Phi,\zeta} : \mbox{Sec}(E)\rightarrow \mbox{Sec}(E)$. The  required result follows from the fact that $T_{\Phi,\zeta}$ is linear. $\mbox{     }\Box$

Given an interaction field $\Phi$ and a gauge $\zeta$ it is natural to consider eigenvectors of the linear operator $T_{\Phi,\zeta}$, such eigenvectors have a distinguished status in Sec$(E)$. If $\psi \in \mbox{Sec}(E)$ then $\psi$ is an eigenvector of $T_{\Phi,\zeta}$ (with positive eigenvalue) if there exists an $m > 0$ such that $T_{\Phi,\zeta}\psi = m\psi$. This is equivalent to the equation
\begin{equation}
\label{eq:Dirac2}
\zeta_{i}(\Sigma^{\mu}\partial_{\mu}-\Phi_{i})\psi_{i} = m\psi_{i}, \forall i\in I.
\end{equation}

\section{Potential fields}

In this section we give a definition of a potential field on $X$. We show how a potential field on $X$ gives rise to an interaction field. A potential field on $X$ may be thought of as a collection of quantities $A_{i\mu}$ such that $(\partial_{\mu}-A_{i\mu})\psi_i$ transforms in a natural way for all $\psi \in \mbox{Sec}(E)$. A natural transformation law is as follows.
\begin{equation}
(\partial_{\mu}-A_{i\mu})\psi_i = (\kappa_{ij}\circ\phi_i^{-1})({{d_{ij}^{-1}}^{\nu}}_{\mu}\circ\phi_i^{-1})((\partial_{\nu}-A_{j\nu})\psi_j)\circ\phi_j\circ\phi_i^{-1}.
\label{eq:potential_transformation1}
\end{equation}
Define a potential field on $X$ to be a collection $A = \{ A_{i\mu} : i\in I, \mu = 0,1,2,3 \}$ of quantities $A_{i\mu} \in \mbox{C}^{\infty}(V_i,\mbox{End}({\bf C}^{4}))$ which have the following transformation properties
\begin{eqnarray}
A_{i\mu}  & = &  {({d_{ij}}^{-1}\circ\phi_i^{-1})^{\nu}}_{\mu} (\kappa_{ij}\circ\phi_{i}^{-1}) (A_{j\nu}\circ (\phi_{j}\circ\phi_{i}^{-1})) (\kappa_{ij}^{-1}\circ\phi_{i}^{-1}) + \nonumber \\
    & & \partial_{\mu}(\kappa_{ij}\circ\phi_{i}^{-1})(\kappa_{ij}^{-1}\circ\phi_{i}^{-1}). \label{eq:formal_potential}
\end{eqnarray} 
\begin{theorem}
Let $A_{i\mu}\in C^{\infty}(V_i,\mbox{End}({\bf C}^4)), i\in I, \mu\in\{0,1,2,3\}$. Then $(\partial_{\mu}-A_{i\mu})\psi_i$ transforms according to Equation~\ref{eq:potential_transformation1} if and only if $A_{i\mu}$ is a potential field.
\end{theorem}
{\bf Proof}
We recall from Equation~\ref{eq:matter_field} that
\begin{equation}
 \psi_{i} = (\kappa_{ij}\circ\phi_{i}^{-1})(\psi_{j}\circ (\phi_{j}\circ\phi_{i}^{-1})), \forall i,j\in I. \nonumber
\end{equation}
Therefore
\[ \partial_{\mu}\psi_i = \partial_{\mu}(\kappa_{ij}\circ\phi_i^{-1})(\psi_j\circ\phi_j\circ\phi_i^{-1})+(\kappa_{ij}\circ\phi_i^{-1})((\partial_{\nu}\psi_j)\circ(\phi_j\circ\phi_i^{-1}))({{d_{ij}^{-1}}^{\nu}}_{\mu}\circ\phi_i^{-1}). \]
Hence the transformation law of Equation~\ref{eq:potential_transformation1} is satisfied if and only if
\[ \partial_{\mu}(\kappa_{ij}\circ\phi_i^{-1})(\psi_j\circ\phi_j\circ\phi_i^{-1})+(\kappa_{ij}\circ\phi_i^{-1})({{d_{ij}^{-1}}^{\nu}}_{\mu}\circ\phi_i^{-1}) - A_{i\mu}(\kappa_{ij}\circ\phi_i^{-1})(\psi_j\circ\phi_j\circ\phi_i^{-1}) \]
\[ = (\kappa_{ij}\circ\phi_i^{-1})({{d_{ij}^{-1}}^{\nu}}_{\mu}\circ\phi_i^{-1})((\partial_{\nu}\psi_j)\circ(\phi_j\circ\phi_i^{-1})-(A_{j\nu}\psi_j)\circ(\phi_j\circ\phi_i^{-1})). \]
Thus, after making a cancellation, we see that transformation law~\ref{eq:potential_transformation1} is satisfied if and only if
\[ \partial_{\mu}(\kappa_{ij}\circ\phi_i^{-1})(\psi_j\circ\phi_j\circ\phi_i^{-1})-A_{i\mu}(\kappa_{ij}\circ\phi_i^{-1})(\psi_j\circ\phi_j\circ\phi_i^{-1}) \]
\[ = -(\kappa_{ij}\circ\phi_i^{-1})({{d_{ij}^{-1}}^{\nu}}_{\mu}\circ\phi_i^{-1})(A_{j\nu}\psi_j)\circ(\phi_j\circ\phi_i^{-1}). \]
This is true for all $\psi_j$ if and only if
\[ \partial_{\mu}(\kappa_{ij}\circ\phi_i^{-1})-A_{i\mu}(\kappa_{ij}\circ\phi_i^{-1}) = -(\kappa_{ij}\circ\phi_i^{-1})({{d_{ij}^{-1}}^{\nu}}_{\mu}\circ\phi_i^{-1})(A_{j\nu}\circ\phi_j\circ\phi_i^{-1}), \]
which is equivalent to the condition that $A_{i\mu}$ be a potential field. $\Box$

In a subsequent paper it will be shown that potential fields for $X$ are induced by linear connections on the total space $TQ$ for the tangent bundle for $Q$.

\begin{theorem}\label{Dirac5}
Suppose that $\{ A_{i\mu} \}$ is a potential field on $X$. Define
\begin{equation}
\Phi_{i} = \Sigma^{\mu} A_{i\mu}, \mbox{ for } i\in I. \label{corr_int_field}
\end{equation}
Then $\{ \Phi_{i} \}$ is an interaction field.
\end{theorem}
{\bf Proof}
Using Equation~\ref{eq:intertwine1} we have,
\begin{eqnarray}
\Phi_{i} & = & \Sigma^{\mu}A_{i\mu} \nonumber \\
    & = & \Sigma^{\mu}(\lambda_{ij}^{-1}\circ\phi_{i}^{-1}){({\Lambda_{ij}}^{-1}\circ\phi_{i}^{-1})^{\nu}}_{\mu}(\kappa_{ij}\circ\phi_{i}^{-1})(A_{j\nu}\circ(\phi_{j}\circ\phi_{i}^{-1})) 
    (\kappa_{ij}^{-1}\circ\phi_{i}^{-1}) + \nonumber \\
    & & \mbox{    } \Sigma^{\mu}\partial_{\mu}(\kappa_{ij}\circ\phi_{i}^{-1})(\kappa_{ij}^{-1}\circ\phi_{i}^{-1}) \nonumber \\
    & = & (\lambda_{ij}^{-1}\circ\phi_{i}^{-1})(\kappa_{ij}\circ\phi_{i}^{-1})\Sigma^{\nu}(A_{j\nu}\circ(\phi_{j}\circ\phi_{i}^{-1}))(\kappa_{ij}^{-1}\circ\phi_{i}^{-1})+ \nonumber \\
    & & \mbox{    } \Sigma^{\mu}\partial_{\mu}(\kappa_{ij}\circ\phi_{i}^{-1})(\kappa_{ij}^{-1}\circ\phi_{i}^{-1}) \nonumber \\
    & = & (\lambda_{ij}^{-1}\circ\phi_{i}^{-1})(\kappa_{ij}\circ\phi_{i}^{-1})(\Phi_{j}\circ(\phi_{j}\circ\phi_{i}^{-1}))(\kappa_{ij}^{-1}\circ\phi_{i}^{-1})+\Sigma^{\mu}\partial_{\mu}(\kappa_{ij}\circ\phi_{i}^{-1}) \nonumber \\
    & & \mbox{    }(\kappa_{ij}^{-1}\circ\phi_{i}^{-1}). \mbox{    }\Box \nonumber
\end{eqnarray}
A special case with $A_{i\mu}$ and $k_{ij}$ commuting is when the potential field components $A_{i\mu}$ are scalar valued quantities. In this case the potential field components transform according to
\begin{equation} \label{potential_commuting}
A_{i\mu} ={ (d_{ij}\circ\phi_{i}^{-1})^{\nu}}_{\mu}(A_{j\nu}\circ (\phi_{j}\circ\phi_{i}^{-1})) + \partial_{\mu}(\kappa_{ij}\circ\phi_{i}^{-1})(\kappa_{ij}^{-1}\circ\phi_{i}^{-1}).
\end{equation} 

\section{A natural intertwining operator between the standard representation of $K$ in Minkowski space and the adjoint representation of $K$ in ${\mathfrak g}$}

In this section we define an intertwining operator between the standard representation of $K$ in Minkowski space and the adjoint representation of $K$ in ${\mathfrak g}$ where ${\mathfrak g}=u(2,2)$ is the Lie algebra of $G=U(2,2)$. We show that this operator is $i$ times the Feynman slash operator with the chiral representation for the Dirac gamma matrices.

Consider $G=U(2,2)$ in the representation in which the metric is given by Eq. \ref{eq:Minkowski}, i.e.
\begin{equation}
g=\left(\begin{array}{cc}
0 & 1 \\
1 & 0
\end{array}\right).
\end{equation}  
Define $\Sigma : u(2) \rightarrow gl(4,{\bf C})$ by
\begin{equation}
\Sigma(M) = 
\left(
\begin{array}{cc}
0 & -M \\
-\pi(M) & 0
\end{array}\right),
\end{equation}
where $\pi$ denotes the parity operator defined by
\begin{equation}
\pi(ix^{\mu}\sigma_{\mu}) = ix^0\sigma_0 - ix^{k}\sigma_{k},
\end{equation}
and $\sigma_{\mu}$ are the Pauli $\sigma$ matrices.
Then, for all $M\in u(2)$
\begin{eqnarray}
\Sigma(M)^{\dagger}g+g\Sigma(M) & = & \left(\begin{array}{cc}
0 & \pi(M) \\
M & 0
\end{array}\right)\left(\begin{array}{cc}
0 & 1 \\
1 & 0
\end{array}\right)+ \\
  & & \left(\begin{array}{cc}
0 & 1 \\
1 & 0
\end{array}\right)\left(
\begin{array}{cc}
0 & -M \\
-\pi(M) & 0
\end{array}\right) \\
    & = & 0
\end{eqnarray}
Also tr$(\Sigma(M)) = 0, \forall M\in u(2)$.
Therefore $\Sigma : u(2) \rightarrow u(2,2)={\mathfrak g}$.

Let $\kappa = \left(\begin{array}{cc} a & 0 \\ 0 & a^{\dagger -1} \end{array} \right) \in K$ where $a\in GL(2,{\bf C}), |\mbox{det}(a)|=1$.
Then for all $M\in u(2)$
\begin{eqnarray}
\kappa\Sigma(M)\kappa^{-1} & = &
 \left(\begin{array}{cc} a & 0 \\ 0 & a^{\dagger -1} \end{array}\right)
 \left(\begin{array}{cc} 0 & -M \\ -\pi(M) & 0 \end{array}\right)
 \left(\begin{array}{cc} a^{-1} & 0 \\ 0 & a^{\dagger} \end{array}\right) \nonumber \\
  & = & \left(\begin{array}{cc} 0 & -aMa^{\dagger} \\ -a^{\dagger -1}\pi(M)a^{-1} & 0 \end{array}\right). \nonumber
\end{eqnarray}
Suppose that 
\[ M = M(x) = ix^{\mu}\sigma_{\mu}. \]
Then
\begin{eqnarray}
\pi(M)M & = & -(x^{0}1-\sum_{i=1}^{3}x^{i}\sigma_{i})(x^{0}1+\sum_{i=1}^{3}x^{i}\sigma_{i}) \nonumber \\
  & = & -(x^{0})^{2}1+\sum_{i,j}x^{i}x^{j}\sigma_{i}\sigma_{j} \nonumber \\
  & = & -Q(M)1, \nonumber 
\end{eqnarray}
where 
\begin{equation}
Q(M) = Q(x) = x^{\mu}x_{\mu} = (x^{0})^{2}-\sum_{i=1}^{3}(x^{i})^{2}.
\end{equation}
Therefore, since det$(M) = -Q(M)$,
\[ \pi(M) = -Q(M)M^{-1}, \]
for $Q(M) \neq 0$. Therefore
\begin{eqnarray}
a^{\dagger -1}\pi(M)a^{-1} & = & a^{\dagger -1}(-Q(M)M^{-1})a^{-1} \nonumber \\
  & = & -Q(M)(aMa^{\dagger})^{-1} \nonumber \\
  & = & -Q(aMa^{\dagger})(aMa^{\dagger})^{-1} \nonumber \\
  & = & \pi(aMa^{\dagger}), \nonumber
\end{eqnarray}
for $Q(M) \neq 0$. It follows by continuity that 
\[ a^{\dagger -1}\pi(M)a^{-1} = \pi(aMa^{\dagger}), \]
for all $M\in u(2)$. Therefore
\begin{equation}
\kappa\Sigma(M(x))\kappa^{-1} = \Sigma(M(\Lambda x)),\forall x\in {\bf R}^{4},
\end{equation}
where $\Lambda$ is the Lorentz transformation corresponding to $\kappa$.
Therefore $\Sigma$ is an intertwining operator between the standard representation of $K$ in ${\bf R}^{4}$ and the adjoint representation of $K$ in ${\mathfrak g}$ with respect to the metric $g$. 

Writing
\begin{equation}
\Sigma(M(x))=ix^{\mu}\gamma_{\mu},
\end{equation}
i.e.
\begin{equation}
i\gamma_{\mu}=\Sigma(M(e_{\mu})),
\end{equation}
where $\{e_{\mu}\}_{\mu=0}^3$ is the standard basis for ${\bf R}^4$, we have that
\begin{equation}
\gamma_0=\left(\begin{array}{cc} 0 & -1_2 \\ -1_2 & 0\end{array}\right), \gamma_j=\left(\begin{array}{cc}0 & -\sigma_j \\ \sigma_j & 0\end{array}\right), \mbox{ for } j=1,2,3,
\end{equation}
and therefore
\begin{equation}
\gamma^0=\left(\begin{array}{cc} 0 & -1_2 \\ -1_2 & 0\end{array}\right), \gamma^j=\left(\begin{array}{cc}0 & \sigma_j \\ -\sigma_j & 0\end{array}\right), \mbox{ for } j=1,2,3,
\end{equation}
and so we recognize $\{\gamma^{\mu}\}_{\mu=0}^3$ to be the Dirac gamma matrices in the chiral representation [23], p. 694. We will call the intertwining operator $\Sigma$ the chiral intertwining operator.

\section{The Dirac equation}

If $\Phi$ is the interaction field induced by a potential field $A$ then the eigenvector Equation~\ref{eq:Dirac2} becomes
\begin{equation} \label{eq:Dirac3}
\zeta_{j}(\Sigma^{\mu}\partial_{\mu}-\Sigma^{\mu}A_{j\mu})\psi_{j} = m\psi_{j},
\forall j\in I.
\end{equation}
With respect to the chiral intertwining operator
 between the fundamental representation of $K$ in ${\bf R}^{4}$ and the adjoint representation of $K$ in ${\mathfrak g}$ Equation~\ref{eq:Dirac3} becomes
\begin{equation}
i\gamma^{\mu}(\partial_{\mu}-A_{j\mu})\psi_{j} = \zeta_j^{-1}m\psi_{j}.
\end{equation}
After making the substitution $A_{j\mu} \rightarrow \frac{e}{i}A_{j\mu}$ this becomes
\begin{equation}
\label{Dirac_equation}
(i{\slas \partial}-e{\slas A}_j)\psi_{j} = \zeta_j^{-1} m\psi_j.
\end{equation}
which is Dirac's equation describing an electron in the presence of an external electromagnetic field with potential $A_{j\mu}$, except that in our case the electron mass $m$ is coupled to gravitation through the gravitational gauge $\zeta$. In the application of the Dirac Equation~\ref{Dirac_equation} to physical problems the quantity $A_j = \{A_{j\mu}\}_{\mu=0}^3$ is the potential field associated with the electromagnetic field.  Define a relative potential for the Dirac equation to be the difference of two potential fields for the Dirac equation.
A relative potential for Dirac's equation has the following transformation property.
\begin{equation}
A_{i\mu} = {({d_{ij}}^{-1}\circ\phi_i^{-1})^{\nu}}_{\mu} (A_{j\nu}\circ (\phi_{j}\circ\phi_{i}^{-1})).
\end{equation} 
Therefore, in other words, it transforms like a covariant vector, or one form.

\section{Gauge invariance}

Gauge invariance is an invariance that theories may possess under certain joint transformations of the potential and the wave function. It is usually considered in flat (Minkowski) space where the potentials transform as 1-forms. In our work the potentials have a more complicated transformation property but we will show that, nevertheless, gauge invariance is manifest. 

Let $\alpha\in C^{\infty}(X,{\bf R})$ be a smooth function and denote $\alpha\circ\phi_i^{-1}$ by $\alpha_i$. Also let $e>0$. Consider the following joint transformation of a potential $\{A_{j\mu}\}$ and a collection of wave functions $\{\psi_j\}$:
\begin{equation}
A_{j\mu}\rightarrow A_{j\mu}+\partial_{\mu}\alpha_{j},
\end{equation}
\begin{equation}
\psi_j\rightarrow (\xi \in V_j\mapsto e^{-ei\alpha_j(\xi)}\psi_j(\xi)), 
\end{equation}
Such a collection of transformations will be called the gauge transformation of \newline $(\{A_{i\mu}\}$,$\{\psi_j\})$ induced by $\alpha$. 
\begin{theorem}
The gauge transformed form of a potential is also a potential.
\end{theorem}
{\bf Proof}
Let $\{A_{i\mu}\}$ be a (scalar) potential. Then
\begin{equation} 
A_{j\mu} = ({{d_{kj}}^{\nu}}_{\mu}\circ\phi_{j}^{-1})(A_{k\nu}\circ (\phi_{k}\circ\phi_{j}^{-1})) + \partial_{\mu}(\kappa_{jk}\circ\phi_{j}^{-1})(\kappa_{jk}\circ\phi_{j}^{-1})^{-1}.
\end{equation}
Thus
\begin{equation}
A_{j\mu}+\partial_{\mu}\alpha_j=({{d_{jk}}^{\nu}}_{\mu}\circ\phi_{j}^{-1})(A_{k\nu}\circ (\phi_{k}\circ\phi_{j}^{-1})) + \partial_{\mu}(\kappa_{jk}\circ\phi_{j}^{-1})(\kappa_{jk}\circ\phi_{j}^{-1})^{-1}+\partial_{\mu}\alpha_j.
\end{equation}
Now
\begin{eqnarray}
\partial_\mu\alpha_j & = & \partial_\mu(\alpha\circ\phi_j^{-1}) \nonumber \\
     &  & \partial_\mu(\alpha\circ\phi_k^{-1}\circ\phi_k\circ\phi_j^{-1}) \nonumber \\
     &  & (\partial_{\nu}\alpha_{k}){{d_{kj}}^{\nu}}_{\mu}\circ\phi_k\circ\phi_j^{-1} \nonumber 
\end{eqnarray}
Therefore
\begin{eqnarray}
A_{j\mu}+\partial_{\mu}\alpha_j  & = & ({{d_{kj}}^{\nu}}_{\mu}\circ\phi_{j}^{-1})(A_{k\nu}\circ (\phi_{k}\circ\phi_{j}^{-1})) + \partial_{\mu}(\kappa_{jk}\circ\phi_{j}^{-1})(\kappa_{jk}\circ\phi_{j}^{-1})^{-1}+ \nonumber \\
    & &   (\partial_{\mu}\alpha_j) \nonumber \\  
  & = & ({{d_{kj}}^{\nu}}_{\mu}\circ\phi_{j}^{-1})((A_{k\nu}+\partial_{\nu}\alpha_k)\circ (\phi_{k}\circ\phi_{j}^{-1})) + \partial_{\mu}(\kappa_{jk}\circ\phi_{j})^{-1} \nonumber \\
    &  & (\kappa_{jk}^{-1}\circ\phi_{j}^{-1}). \nonumber 
\end{eqnarray}
Thus $\{A_{j\mu}+\partial_{\mu}\alpha_j\}$ transforms as a potential as required.
 $\Box$

One can easily show that the gauge transformation of a section $\psi=\{\psi_i\}\in\mbox{Sec}(E)$ is an element of Sec$(E)$. It is now straightforward to prove the following.
\begin{theorem}
Let $\alpha\in C^\infty(X,{\bf R}), \{A_{j\mu}\}$ a (scalar) potential for $X$ and $\{\psi_j\}$ a solution to Dirac's equation relative to $\{A_{j\mu}\}$. Then the gauge transformations of $\{A_{j\mu}\}$ and $\{\psi_j\}$ together satisfy Dirac's equation.
\end{theorem}

\section{Reference fields and relative fields}

In this section we show that the space of interaction fields can be represented in terms of a certain operator algebra bundle. 

Let $F$ be the associated bundle to $Q$ associated with the adjoint representation of $K$ in End$({\bf C}^4)$. Since the representation is by inner automorphisms, $F$ has the structure of a bundle of algebras. Let Sec$(F)$ denote the space of smooth sections of $F$. A collection of quantities $\Psi =
 \{ \Psi_{i} \}_{i\in I}$, where $\Psi_{i}\in \mbox{C}^{\infty}(V_i,{\bf C}^{4\times 4}), \forall i\in I$, is an element of Sec$(F)$ if and only if it has the following transformation property.
\begin{equation}
\Psi_{i} = (\kappa_{ij}\circ\phi_{i}^{-1})(\Psi_{j}\circ (\phi_{j}\circ\phi_{i}^{-1}))(\kappa_{ij}\circ\phi_{i}^{-1})^{-1}.
\end{equation}
Elements of Sec$(F)$ will be called relative fields.

If $\zeta \in {\cal G}$ let $\zeta^{-1}$ denote the collection of quantities $\{\zeta_i^{-1}\}$ (multiplicative inverse). If $\Theta = \{\Theta_i\}$ is a collection of quantities $\Theta_i : V_i\rightarrow Y$, for some vector space $Y$, then let $\zeta^{-1}\Theta$ denote the collection of quantities defined by
\[ (\zeta^{-1}\Theta)_i = \zeta_i^{-1}\Theta_i, \]
with pointwise multiplication over $V_i$.
It is clear that if $\Phi_{1}, \Phi_{2} \in {\cal F}$ and $\zeta\in {\cal G}$ then there exists $\Psi\in \mbox{Sec}(F)$ such that $\Phi_{1}-\Phi_{2} = \zeta^{-1}\Psi$. Also, if $\Phi\in{\cal F}, \Psi\in\mbox{Sec}(F)$ and $\zeta\in{\cal G}$ then $\Phi+\zeta^{-1}\Psi\in{\cal F}$.
From this follows the following theorem.
\begin{theorem} \label{Dirac4}
Given any interaction field $\Phi^{(0)}$ as a reference field and a gauge $\zeta\in {\cal G}$, the space of all interaction fields can be written as
\begin{equation}
{\cal F} = \{ \Phi^{(0)}+\zeta^{-1}\Psi:\Psi\in\mbox{Sec}(F)\}.
\end{equation}
\end{theorem}
This shows the affine nature of the space of interaction fields.

\section{Gravitation \label{section:gravitation}}

Tangent vectors transform contravariantly under the group $J$ (see Eq. \ref{eq:transformation_derivative}). 
Suppose that $\zeta = \{ \zeta_{i} \}$ is a gauge for $X$. Then by Equation~\ref{eq:gauge}
\begin{equation}
\zeta_{i}(\phi_{i}(x)) = \lambda_{ij}(x)\zeta_{j}(\phi_{j}(x)),
\end{equation}
for $x\in U_i\cap U_j$ and $i,j\in I$. Given $\zeta$ we can define a metric on $X$ as follows. For $x\in X$ and $u,v\in T_{x}X$ define 
\begin{equation}
(u,v) = \zeta_{i}(\phi_{i}(x))^{-2}(u_{i},v_{i})_M,
\end{equation}
where $($ $,$ $)_M : {\bf R}^{4}\times{\bf R}^{4} \rightarrow {\bf R}$ is the Minkowski space metric, $i\in I$ and $u_i$ and $v_i$ are the values of $u$ and $v$ in coordinate system $i$. Now
\begin{eqnarray}
\zeta_i(\phi_i(x))^{-2}(u_i,v_i)_M & = & (\lambda_{ij}(x)\zeta_j(\phi_j(x)))^{-2}(d_{ij}u_j,d_{ij}v_j)_M \nonumber \\
  & = & (\lambda_{ij}(x)\zeta_{j}(\phi_{j}(x)))^{-2}(\lambda_{ij}(x)\Lambda_{ij}(x)u_{j},\lambda_{ij}(x)\Lambda_{ij}(x)v_{j})_M \nonumber \\
  & = & \zeta_{j}(\phi_{j}(x))^{-2}(u_{j},v_{j})_M, \nonumber
\end{eqnarray}
$\forall x\in X, u,v\in T_xX,i,j\in I$. This shows that the definition of the inner product in $T_{x}X$ is invariant under a change of coordinate system and therefore that the inner product is well defined. The metric for $X$ defined in this way will be called the standard metric associated with the given gauge. 

We consider gravitation to be associated with the geometry and/or topology of spacetime providing a background for the fundamental interactions such as the electroweak interaction. This approach is in the spirit of Einstein's general theory of relativity. In our case the space-time is locally conformally flat and the physics is conformally invariant (which may be compared with Weyl's approach). 

Further work needs to be done on this approach to gravity in regards to the experimental tests of general relativity involving the non- locally conformally flat exterior Schwarzschild metric (vacuum solution), though the interior Schwarzschild metric is known to be conformally flat [24]. Such work may relate to the recently discussed AdS/Ricci-flat correspondence or the Schwarzschild solution on the brane [25]. Also, as is well known,  ``empty space" in QFT is full of virtual particles whch may emerge in further development of this work. 

Mannheim [26] considered an action for the Universe of conformal form involving the Weyl conformal tensor. Functional variation of the action with respect to the metric leads to the equation $R_{\mu\nu}=0$ as a vacuum solution, where $R_{\mu\nu}$ is the Ricci tensor, leading therefore to the Schwarzschild solution. This shows that the Einstein gravitation equations are sufficient but not necessary to give rise to the Schwarzschild solution. 

In standard approaches to quantum gravity, gravitation seems to be at odds with quantum theory.
It is known that conventional QFT techniques fail when applied to gravitation.

Quantum gravity in the context of conformally flat space-times has been studied by a number of authors (e.g. Hamada, [27]).

The main advantage of considering locally conformally flat space-times, in fact M\"{o}bius structures, is the structure of the differential invariants of their associated vector bundles leading to, in this paper, an {\em ab initio} derivation of Dirac's equation for the electron and also, shortly, to the derivation of Maxwell's equations. 

It will also be shown in a subsequent paper that when one considers a certain unitary representation of $K$ on an infinite dimensional topological vector space quantum electrodynamics (QED) emerges in a natural way through the application of natural invariance principles.   

\section{The cohomology of alternating multilinear fields and Maxwell's equations}

In this section we give a derivation of the vacuum Maxwell equations by considering the standard and canonical differential forms representing an alternating multilinear field. Distinguished alternating multilinear fields are identified by  means of the de Rham cohomology applied to the space of differential forms providing a canonical representation for the alternating multilinear fields.

The canonical representation for an alternating multilinear field is found to be obtained from the standard representation by means of the Hodge star operator. This operator will be defined if there is present a metric on the manifold. We know from Section \ref{section:gravitation} that given any gauge for a M\"{o}bius structure $X$ there is induced a natural metric on $X$.

\subsection{Alternating multilinear fields}

Let Alt$_n(X)$ denote the space of sections of the bundle $\bigcup_{x\in X} \mbox{Alt}_n(T_x X,{\bf R})$. Elements of Alt$_n(X)$ can be thought of as smooth maps $\Psi:x\in X\mapsto\Psi(x)$  where $\Psi(x)\in\mbox{Alt}(T_x(X),{\bf R}),\forall x\in X$.
Elements of Alt$_n(X)$ will be called alternating multilinear fields.
Consider the action $\rho: J\times\mbox{Alt}_n({\bf R}^4,{\bf R})\rightarrow \mbox{Alt}_n({\bf R}^4,{\bf R})$ defined by
\begin{equation}
\rho(\Lambda,\Psi)(v_1,\ldots,v_n) = \Psi(\Lambda^{-1}v_1,\ldots,\Lambda^{-1}v_n).
\end{equation}
Then Alt$_n(X)$ is the vector bundle associated through this action to the principal bundle $R$. 

\subsection{The standard differential form representing an alternating multilinear field}

There is a canonical isomorphism between Alt$_{n}(T_{x}X,{\bf R})$ and $(\bigwedge_{n}(T_{x}X))^{*}$ [28], p. 58. Suppose that we have a  non-singular pairing $($ $,$ $) : \bigwedge_{n}(T_{x}^{*}X)\times \bigwedge_{n}(T_{x}X)\rightarrow{\bf R}$. An example of such a pairing is the standard pairing defined by
\begin{equation}
(u^{*},v) = (u^{*},v)_S = \frac{1}{n!} \det(u_{i}^{*}(v_{j})|_{i,j=1,\ldots,n}),
\end{equation}
where $u^{*} = u_{1}^{*}\wedge \cdots \wedge u_{n}^{*}$ and $v = v_{1} \wedge \cdots \wedge v_{n}$. Such a pairing gives rise to an isomorphism of $(\bigwedge_{n}(T_{x}X))^{*}$ with $\bigwedge_{n}(T_{x}^{*}X)$ and therefore, given the previously mentioned canonical isomorphism, to an isomorphism of Alt$_{n}(T_{x}X,{\bf R})$ with $\bigwedge_{n}(T_{x}^{*}X)$.

Thus, given an alternating multilinear field $\Psi$ and a choice of pairing $(\mbox{ , })$ we have, using the universal mapping property of $\wedge$ that there is a unique differential form $\omega(\Psi)\in\bigwedge_n(T_x^{*}X)$ such that
\begin{equation}
\label{Yang1}
(\Psi(x))(v_{1}, \ldots , v_{n}) = ((\omega(\Psi))(x),v_{1}\wedge \cdots \wedge v_{n}), \forall x\in X, v_{1}, \ldots , v_{n}\in T_{x}X.
\end{equation}

If the standard pairing is used then we call $\omega(\Psi)$ the standard differential form representing $\Psi$. However, in general, $\omega(\Psi)$ is pairing dependant and hence not canonical.

\subsection{The canonical differential form representing  an alternating multilinear field}

The inner product in $T_{x}X$ induced by a gauge induces, in a canonical way, an isomorphism between $T_{x}X$ and $T_{x}^{*}X$ which we may denote by $v\mapsto v^{*}$. This, given a pairing $(\mbox{ },\mbox{ }):\bigwedge_n(T_x^{*}X)\times\bigwedge_n(T_x X)\rightarrow{\bf R}$,  induces an inner product on $\bigwedge_n(T_x^{*}X)$ according to
\begin{equation}
(u_1^{*}\wedge\ldots\wedge u_n^{*},v_1^{*}\wedge\ldots\wedge v_n^{*})=(u_1^{*}\wedge\ldots\wedge u_n^{*},v_1,\ldots, v_n).
\end{equation}
It follows from Eq. \ref{Yang1} 
\begin{equation}
(\Psi(x))(v_1,\ldots,v_n)=((\omega(\Psi))(x),v_1^{*}\wedge\ldots\wedge v_n^{*}), \forall v_1,\ldots, v_n\in T_x X.
\end{equation}
Also $\bigwedge_{4}(T_{x}^{*}X)$ can be canonically identified with ${\bf R}$ by means of the volume form $\tau$. Therefore, with respect to any given metric, there is a canonical map
${\ch \Psi}(x): \bigwedge_{n}(T_{x}^{*}X) \rightarrow \bigwedge_{4}(T_{x}^{*}X)$, satisfying
\begin{equation}
({\ch \Psi}(x))( v_{1}^{*}\wedge \cdots \wedge v_{n}^{*})=(\omega(\Psi)(x),v_1^{*}\wedge\ldots\wedge v_n^{*})\tau(x)=(\Psi(x)( v_{1}, \ldots , v_{n})) \tau(x), 
\end{equation}
$\forall x\in X$ and $v_{1}, \ldots , v_{n}\in T_xX$.
${\ch \Psi}$ is pairing independent but metric dependent. Now we recall [29], p. 295, that the star operator $* : \bigwedge_n(T^*_xX)\rightarrow\bigwedge_{4-n}(T^*_xX)$ where $n\in\{0,1,2,3\}$ is the isomorphism defined by
\begin{equation}
(\alpha,\beta)\tau = \alpha\wedge *\beta,
\end{equation}
for all $\alpha, \beta \in \bigwedge_n(T_x^*X)$,
where $(\mbox{ , })$ is the metric on $\bigwedge_n(T_x^*X)$ induced by the metric on $T_xX$ and the given pairing. Thus
\begin{eqnarray}
({\ch \Psi}(x))( v_{1}^{*}\wedge \cdots \wedge v_{n}^{*}) & = & ((\omega(\Psi))(x), v_{1}^{*}\wedge \cdots \wedge v_{n}^{*})\tau(x) \nonumber \\
    & = & (v_{1}^{*}\wedge \cdots \wedge v_{n}^{*},(\omega(\Psi))(x))\tau(x) \nonumber \\
    & = & v_{1}^{*}\wedge \cdots \wedge v_{n}^{*} \wedge *(\omega(\Psi))(x) \nonumber \\
    & = & (-1)^{n(4-n)} *(\omega(\Psi))(x)\wedge v_{1}^{*} \wedge \cdots \wedge v_{n}^{*} \nonumber \\
    & = &  (-1)^{n(4-n)}*(\omega(\Psi))(x)\wedge(v_{1}^{*}\wedge \cdots \wedge v_{n}^{*}), \nonumber
\end{eqnarray}
and so, through the left regular representation of $\bigwedge(T_{x}^{*}X)$ considering elements of $\bigwedge(T_{x}^{*}X)$ to be operators on $\bigwedge(T_{x}^{*}X)$, we may write
\begin{equation}
{\ch \Psi}(x) = (-1)^{n(4-n)}*(\omega(\Psi))(x).
\end{equation}
and therefore
\begin{equation}
{\ch \Psi} = (-1)^{n(4-n)}*\omega(\Psi).
\end{equation}
${\ch \Psi}$ is metric dependent and therefore may be described as being gauge dependent.
Since ${\ch \Psi}$ is a pairing independent but gauge dependent representation of $\Psi$ it follows that $(-1)^{n(4-n)}*\omega(\Psi)$ is a  pairing independent but gauge dependent, representation of $\Psi$.

In four dimensions two metrics on a smooth manifold which are conformally related define the same Hodge * operator on two forms [30]. Therefore the Hodge * operator on two forms is gauge independent.

\subsection{Maxwell's equations}

We will now consider whether there are distinguished alternating multilinear fields. It would then be reasonable to consider that these distinguished fields have physical significance.

We have seen how any alternating multilinear field $\Psi\in\mbox{Alt}_{n}(X)$ is associated with its standard representation $\omega(\Psi)$ as a differential form and a canonical representation $(-1)^{n(4-n)}*\omega(\Psi)$. We have a natural way of identifying distinguished differential forms by means of the de Rham cohomology theory. Representatives of the cohomology modules are the closed forms. 

For practical representation of alternating multilinear fields we use the standard representation. However the canonical representation is relevant for seeking distinguished fields.
It is expected that a canonical representation of a space will have the same structure as the space and will not have structure due to arbitrary choices made. Therefore in seeking distinguished elements of a space through a representation of the space the canonical representation should be used so that elements are distinguished relative to the actual structure of the space itself.
Therefore, we seek distinguished two forms $F = \omega(\Psi)$ according to
\begin{equation}
d(-1)^{n(4-n)}*F = 0,
\end{equation}
where $d$ is the exterior derivative operator. This is equivalent to 
\begin{equation}
\label{eq:YangMills}
d*F = 0,
\end{equation} 
which is of the form of the Yang-Mills vacuum field equation.

Let $A_{j\mu}$ be a potential for the Dirac equation. Suppose that $\phi_j\in {\mathcal A}$ is any coordinate system and consider all coordinate systems $\phi_{j{\prime}}$ which are related to $\phi_j$ by a Poincare transformation, i.e.
\[ U_j\cap U_j^{\prime}\neq\emptyset \mbox{ and }\exists \Lambda\in O(1,3)^{\uparrow+}, c\in{\bf R}^4,  (\phi_{j^\prime}\circ\phi_j^{-1})(\xi)=\Lambda\xi+c, \forall\xi\in\phi_j(U_j\cap U_{j^\prime}). \]
Let ${\mathcal P}_j$ denote the collection of all such coordinate systems. Then $\forall\phi_{j^{\prime}}\in{\mathcal P}_j, \kappa_{j^{\prime}j}$ is constant and so $(\partial_\mu(\kappa_{{j^{\prime}j}}\circ\phi_{j^{\prime}}^{-1}))(\xi)=0, \forall\xi\in\phi_{j^{\prime}}(U_j\cap U_{j^{\prime}})$. Thus by Eq. \ref{potential_commuting}
\begin{equation}
A_{j^{\prime}\mu}={(d_{j^{\prime}j}\circ\phi_{j^{\prime}}^{-1})^{\nu}}_{\mu}(A_{j\nu}\circ(\phi_{j}\circ\phi_{j^{\prime}}^{-1})), \forall\phi_{j^{\prime}}\in{\mathcal P}_j.
\end{equation}
Thus $A_{j^{\prime}}$ transforms as a 1-form between coordinate systems in ${\mathcal P}_j$.

Let $F = dA$. Then the condition that $F$ satisfy Equation~\ref{eq:YangMills} is equivalent to the vacuum Maxwell's equations.
If Maxwell's equations are considered on spaces with non-trivial topology such as ${\bf R}^4$ with ``wormholes'' then the free space field acquires the properties of a field in the presence of charges [29].

\section{Conclusion}

We have shown that classical field theory and $(1^{\mbox{st}}$ quantized) quantum mechanics as described by Maxwell's and Dirac's equations can be derived by considering the bundle $Q$ associated with any M\"{o}bius structure through representations of its structure group on ${\bf C}^4$ and ${\bf R}^4$ respectively. An advantage of such a formulation is that it can be readily axiomatized using only a few natural axioms involving causal structure and mathematical naturalness. Furthermore it generalizes classical quantum field theory from the context of Minkowski space to that of arbitrary locally conformally flat space-times. 

In the usual development of physics, principles represented by field equations such as Maxwell's equations are postulated or presented because they successfully describe the experimentally observed data. When the equations are derived from variational principles objects such as Lagrangian densities are postulated because they can be used to derive the given field equations. Dirac's equation is usually derived by seeking a Lorentz invariant first order linear partial differential equation whose solutions satisfiy the Klein-Gordon equation, i.e. are eigenfunctions of the wave operator. The great importance of the wave operator in physics is due to its success in describing electromagnetic and other fields, it is not derived as a product of any underlying principles (the ansatz $p_{\mu}\rightarrow i\partial_{\mu}-eA_{\mu}$ is also justified by its success).

The most important reason for formulating field theory using the principal bundle $Q$ with structure group $K$ is, as will be shown in a subsequent paper, that QED can be naturally formulated in terms of a bundle associated with a certain infinite dimensional unitary representation of $K$. 

\section*{Acknowledgements} 
The work described in this paper was supported by Melbourne University, Stony Brook University and the Commonwealth Scientific and Industrial Research Organisation (CSIRO, Australia). The author is particularly grateful to Hyam Rubinstein and Sergei Kuzenko for supporting this work and to Iain Aitchison for very helpful discussions.  

\section*{References}

\rf [1] Mashford, J. S., ``A non-manifold theory of space-time", {\em Journal of Mathematical Physics} 22(9), 1981, 1990-1993.

\rf [2] Mashford, J. S., {\em Invariant measures and M\"{o}bius structures: A framework for field theory}, PhD thesis, University of Melbourne, 2005.

\rf [3] Kobayashi, S. and Nomizu, K., {\em Foundations of Differential Geometry}, Volume I, Wiley, New York, 1963.

\rf  [4] Kulkarni, R. S., ``Conformal structures and M\"{o}bius structures", In: {\em Conformal Geometry},  Kulkarni, R. S. and Pinkall, U. (eds.), Max-Planck-Institut f\"{u}r Mathematik, Bonn, 1988.

\rf [5] Thurston, W., {\em Geometry and topology of 3-manifolds}, Princeton, 1978.

\rf [6] Bestvina, M., ``Geometric group theory and 3-manifolds hand in hand: the fulfillment of Thurston's vision", {\em Bulletin of the American Mathematical Society} 51(1), 2014, 53-70.

\rf [7] Dirac, P.A.M., ``Wave equations in conformal space", {\em Annals of Mathematics}, Second Series, Vol. 37, No. 2, 1936, 429-442.

\rf [8] Cartan, \'{E}., ``Sur les vari\'{e}t\'{e}s \`{a} connexion affine et la th\'{e}orie de la relativit\'{e} g\'{e}n\'{e}ralis\'{e}e (premi\`{e}re partie)", {\em Annales Scientifiques de l'\'{E}cole Normale Sup\'{e}rieure} 40, 1923, 325-–412.

\rf [9] Brozos-Vazquez, M., Garcia-Rio, E., Vazquez-Lorenzo, R., ``Some remarks on locally conformally flat static space-times", {\em Journal of Mathematical Physics}, 46(2), 2005.

\rf [10] Cabral, F. and Lobo, F.S.N., ``Electrodynamics and spacetime geometry: Foundations", {\em Foundations of Physics} 47(2), 2017, 208-228.

\rf [11] Schwarz, A., ``Axiomatic conformal theory in dimension greater than 2 and the AdS/CT Correspondence", {\em  Letters in Mathematical Physics}, 2016, 1181-1197.

\rf [12] Schoen, R. and Yau, S.-T., ``Conformally flat manifolds, Kleinian groups and scalar curvature", {\em Inventiones Mathematicae} 92, 1988, 47-71.

\rf [13] Izeki, H., ``Limit set of Kleinian groups and conformally flat Riemannian manifolds", {\em Inventiones Mathematicae} 122, 1995, 603-625.

\rf [14] Kastrup, H. A., ``On the advancements of conformal transformations and their associated symmetries in geometry and theoretical physics", {\em Ann. Phys.} (Berlin) 17(9-10), 2008, 631-690.

\rf [15] Penrose, R., ``Twistor theory and the Einstein vacuum", {\em Classical and Quantum Gravity} A 16, 1999, 113-130.

\rf [16] 't Hooft, G., ``A class of elementary particle models without any adjustable real parameters", {\em Foundations of Physics} 41, 2011, 1829-1856.

\rf [17] Carlip, S., ``Quantum gravity: A progress report". {\em Reports on Progress in Physics} 64(8), 2001, 885-942.

\rf [18] Bombelli, L., Lee, J., Meyer, D. and Sorkin, R. D., ``Space-time as a causal set". {\em Physical Review Letters} 59(5), 1987, 521-524.

\rf [19] Doebner. H. D., Hennig, J. D. and Palev, T. D., {\em Supermanifolds, Geometrical Methods and Conformal Groups}, World Scientific, 1989.

\rf [20] Kosmann, Y., ``D\'{e}riv\'{e}es de Lie des spineurs", {\em Annali di Matematica Pura ed Applicata} 91(4), 1972, 317-395.

\rf [21] Michel, J.-P., ``Conformal geometry of the supercotangent and spinor bundles", {\em Communications in Mathematical Physics} 312, 2012, 303-330.

\rf  [22] Eastwood, M. G. and Rice, J. W., ``Conformally invariant differential operators on Minkowski space and their curved analogues", {\em Comm. Math. Phys.} 109(2), 1987, 207-28.

\rf [23] Itzykson, C. and Zuber, J.-B., {\em Quantum Field Theory}, McGraw Hill, 1980.

\rf [24] Herrero, A. and Morales, A., ``Schwarzschild interior in conformally flat form", {\em General Relativity and Gravitation}, Vol. 36., No. 9., 2004.

\rf [25] Chakraborty, S. and Bandyopadhyay, T., ``Schwarzschild solution on the brane", {\em International Journal of Theoretical Physics} 47, 2008, 2493-2499.

\rf [26] Mannheim, P. D., ``Making the case for conformal gravity", {\em Foundations of Physics} 42, 2012, 388-420.

\rf [27] Hamada, K.-j., ``Background-free quantum gravity based on conformal gravity and conformal field theory on $M^4$", {\em Physical Review D}, 85(2), 2012.

\rf [28] Warner, F. W., {\em Foundations of Differentiable Manifolds and Lie Groups}, Springer Verlag, New York, 1983.

\rf [29] Choquet-Bruhat, Y., DeWitt-Morette, C. and Dillard-Bleick, M., {\em Analysis, Manifolds and Physics}, North-Holland, Amsterdam, 1982.

\rf [30] Dray, T., Kulkarni, R. and Samuel, J., ``Duality and conformal structure", {\em Journal of Mathematical Physics} 30, 1989, 1306-1309.

\end{document}